\RequirePackage{filecontents}

\RequirePackage{snapshot}
\documentclass[aps,prl,reprint,showpacs,secnumarabic]{preprint}

\renewcommand\documentclass[2][]{}
\let\origparagraph=\paragraph
\AtBeginDocument{%
  \let\paragraph=\origparagraph
}
\makeatother
\documentclass[reprint,aps,prl,showpacs,floatfix]{revtex4-1}

\usepackage[T1]{fontenc}
\usepackage{amsmath}
\usepackage{txfonts}
\usepackage{tgtermes}
\usepackage[altg]{qtxmath}
\renewcommand*{\vec}[1]{\boldsymbol{#1}}
\usepackage{microtype}
\usepackage{graphicx}
\usepackage{braket}
\usepackage{mdwlist}
\graphicspath{{figures/}}
\usepackage[breaklinks=true,
pdfauthor={Nicolas Teeny, Enderalp Yakaboylu, Heiko Bauke and Christoph H. Keitel},
pdftitle={Ionization Time and Exit Momentum in Strong-Field Tunnel Ionization}]{hyperref}
\newcommand*{\D}{\mathrm{d}}

\newcommand*{\I}{\mathrm{i}}

\DeclareMathOperator*{\argmax}{arg\,max}

\makeatletter
\def\paragraph#1{%
  \@startsection
    {paragraph}%
    {4}%
    {\parindent}%
    {\z@}%
    {-1sp}%
    {\normalfont\normalsize\itshape}{#1.---}%
}%
\makeatother

\begin{document}

\title{Ionization Time and Exit Momentum in Strong-Field Tunnel Ionization}

\author{Nicolas Teeny}
\affiliation{Max-Planck-Institut f{\"u}r Kernphysik, Saupfercheckweg~1, 69117~Heidelberg, Germany}
\author{Enderalp Yakaboylu}
\affiliation{Max-Planck-Institut f{\"u}r Kernphysik, Saupfercheckweg~1, 69117~Heidelberg, Germany}
\author{Heiko Bauke}
\affiliation{Max-Planck-Institut f{\"u}r Kernphysik, Saupfercheckweg~1, 69117~Heidelberg, Germany}
\author{Christoph H. Keitel}
\affiliation{Max-Planck-Institut f{\"u}r Kernphysik, Saupfercheckweg~1, 69117~Heidelberg, Germany}

\begin{abstract}
  Tunnel ionization belongs to the fundamental processes of atomic
  physics. The so-called two-step model, which describes the ionization
  as instantaneous tunneling at the electric field maximum and classical
  motion afterwards with zero exit momentum, is commonly employed to
  describe tunnel ionization in adiabatic regimes.  In this
  contribution, we show by solving numerically the time-dependent
  Schr{\"o}dinger equation in one dimension and employing a virtual detector
  at the tunnel exit that there is a nonvanishing positive time delay
  between the electric field maximum and the instant of ionization.
  Moreover, we find a nonzero exit momentum in the direction of the
  electric field.  To extract proper tunneling times from asymptotic
  momentum distributions of ionized electrons, it is essential to
  incorporate the electron's initial momentum in the direction of the
  external electric field.
\end{abstract}

% 32.80.Rm 	Multiphoton ionization and excitation to highly excited states
% 31.30.J- 	Relativistic and quantum electrodynamic (QED) effects in atoms, molecules, and ions
% 03.65.Xp 	Tunneling, traversal time, quantum Zeno dynamics
% 32.80.Fb 	Photoionization of atoms and ions 

\date{February 10, 2016}

\maketitle

%\paragraph{Introduction}

\begin{figure}[b]
  \centering
  \includegraphics[scale=0.45]{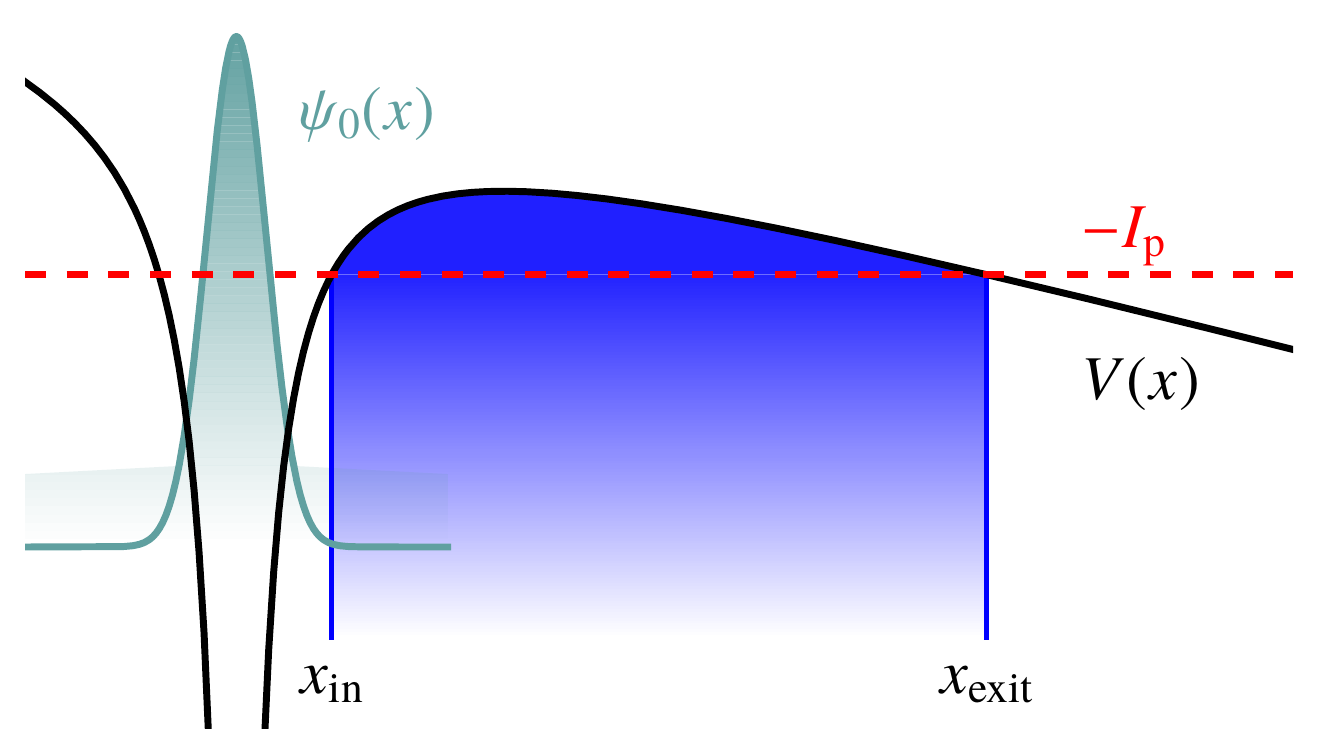}
  \caption{%(color online)
    The one-dimensional Coulomb potential, which is bent by an
    additional electric field, defines the effective potential $V(x)$
    and the tunneling barrier between the points $x_\mathrm{in}$ and
    $x_\mathrm{exit}$ for an electron, which is given by the wave
    function $\psi_0(x)$, with the ground state energy $-I_\mathrm{p}$.}
  \label{fig:barrier}
\end{figure}

In 1932, MacColl \cite{MacColl:1932:Note_on} studied the time which may
be associated with the process of a particle approaching from far away a
potential barrier of a height larger than the particle's energy and
eventually tunneling through the barrier. Many efforts have been
directed toward defining
\cite{Hauge:Stvneng:1989:Tunneling_times,Landauer:Martin:1994:Barrier_interaction,Maji:Mondal:etal:2007:Tunnelling_time}
and measuring
\cite{Steinberg:Kwiat:etal:1993:Measurement_of,Gueret:Marclay:Meier:1988,Mugnai:Ranfagni:etal:2000:Observation_of,Enders:Nimtz:1993:Evanescent-mode_propagation,Winful:2006:Tunneling_time}
tunneling times.  A related open problem is the question of how long it
takes to ionize by tunneling through a binding potential.  Keller
\textit{et~al.}\ have conducted experiments using the angular streaking
technique aiming to measure tunneling times for ionization from a bound
state, the so-called attoclock experiments
\cite{Eckle:Pfeiffer:etal:2008:Attosecond_Ionization,Eckle:Smolarski:etal:2008:Attosecond_angular}.
In the tunnel ionization case, a Coulomb-bound electron is ionized by a
strong electromagnetic field, and a potential barrier can be defined via
the electron's binding energy and the Coulomb potential bent by the
electric field's potential; see Fig.~\ref{fig:barrier}.  Since the
attoclock experiments have been performed, many renewed efforts have
been directed toward defining a tunnel ionization time, because a
consensus on a suitable theoretical definition of tunneling time and the
interpretation of experimental results is still lacking
\cite{keldysh,Maquet:Caillat:etal:2014:Attosecond_delays,Eckle:Smolarski:etal:2008:Attosecond_angular,Eckle:Pfeiffer:etal:2008:Attosecond_Ionization,Orlando:McDonald:etal:2014:Tunnelling_time,Klaiber:Yakaboylu:etal:2013:Under-the-Barrier_Dynamics,Landsman:Keller:2015:Attosecond_science,Zhao:Lein:2013:Determination_of}.

In the following, we study the time delay $\tau_A$ between the instant
of ionization, i.\,e., when the electron exits the barrier, and the
instant of electric field maximum $t_0$.  One may reconstruct the moment
of ionization from the electron's asymptotic momentum $\vec p(\infty)$,
which reads with the electron's charge $q$ and the time-dependent
electric field $\vec E(t)$ (applying the dipole approximation and
neglecting Coulomb corrections)
\begin{equation}
  \label{eq:pend}
  \vec p(\infty) = 
  \vec p(t_0+\tau_\mathrm{A}) + 
  q \int_{t_0+\tau_\mathrm{A}}^\infty \vec E(t')\,\D t'\,.
\end{equation}
In attoclock experiments, electrons are ionized by elliptically
polarized light, which makes the direction of the asymptotic momentum
very sensitive to the ionization time \mbox{$t_0+\tau_\mathrm{A}$}.  In
the extraction of $\tau_\mathrm{A}$ from attoclock experiments, it is
state of the art to treat the ionized electron classically, to take into
account Coulomb corrections, and to assume that the electron's initial
momentum follows from some semiclassical theory
\cite{Landsman:Weger:etal:2014:Ultrafast_resolution,Delone:Krainov:1991:Energy_and}.
For a reliable reconstruction of the attoclock time $\tau_\mathrm{A}$,
however, suitable initial conditions have to be identified as pointed
out in
Refs.~\cite{Kaushal:Smirnova:2013:Nonadiabatic_Coulomb,Torlina:Morales:etal:2015:Interpreting_attoclock}.
In particular, assumptions about the initial momentum bias the
reconstructed value for $\tau_\mathrm{A}$.

The popular two-step model of tunnel ionization assumes a maximal
ionization rate at the instant of maximal electric field strength,
i.\,e., $\tau_\mathrm{A}=0$, and that free electrons have zero initial
momentum, i.\,e., $\vec p(t_0+\tau_\mathrm{A})=0$.  Within the two-step
model, the electron's asymptotic momentum follows by solving the
classical equations of motion for the electron's motion in the combined
electromagnetic field of the binding potential and the ionizing external
light pulse.  This model, however, cannot be used as a benchmark for
experiments, because a possible match or mismatch of experimental data
and the theoretical prediction by the two-step model may be explained by
various pairs of nonzero delay $\tau_\mathrm{A}$
\cite{Lein:2011:Streaking_analysis} and nonzero initial momentum $\vec
p(t_0+\tau_\mathrm{A})$, which may a have nonzero component parallel to
the electric field direction
\cite{Pfeiffer:Cirelli:etal:2012:Probing_Longitudinal,Yakaboylu:Klaiber:etal:2014:Wigner_time}.

Neither the delay $\tau_\mathrm{A}$ nor the initial momentum $\vec
p(t_0+\tau_\mathrm{A})$ are directly accessible by experiments, and it
is also challenging to calculate them analytically.  Therefore, we
employ \textit{ab~initio} quantum calculations and a virtual detector
\cite{Wang:Tian:etal:2013:Extended_Virtual,Feuerstein:Thumm:2003:On_computation}
at the tunnel exit.  The virtual detector technique allows us to
determine directly the electron's time of arrival at the tunnel exit as
well as its exit momentum.

We analyze theoretically an initially bound electron ionized by an
electric field pulse.  To determine the time delay $\tau_\mathrm{A}$ and
the exit momentum, we solve the time-dependent Schr{\"o}dinger equation and
place a virtual detector at the tunneling exit.  The virtual detector is
realized by calculating the probability current at the exit.  The most
probable time delay $\tau_\mathrm{A}$ is determined by comparing the
instant of maximum probability current at the potential barrier exit and
the instant of maximum electric field strength.  The exit momentum is
determined by the space-resolved momentum distribution at the tunnel
exit at the instant of ionization.  By separating the wave function into
a tunneled part and a bound part after the interaction with the laser
pulse, we can calculate the momentum distribution of the tunneled
electron, from which one can determine the most probable asymptotic
momentum.

\paragraph{Considered system}
   
\begin{figure}[b]
  \centering
  \includegraphics[scale=0.45]{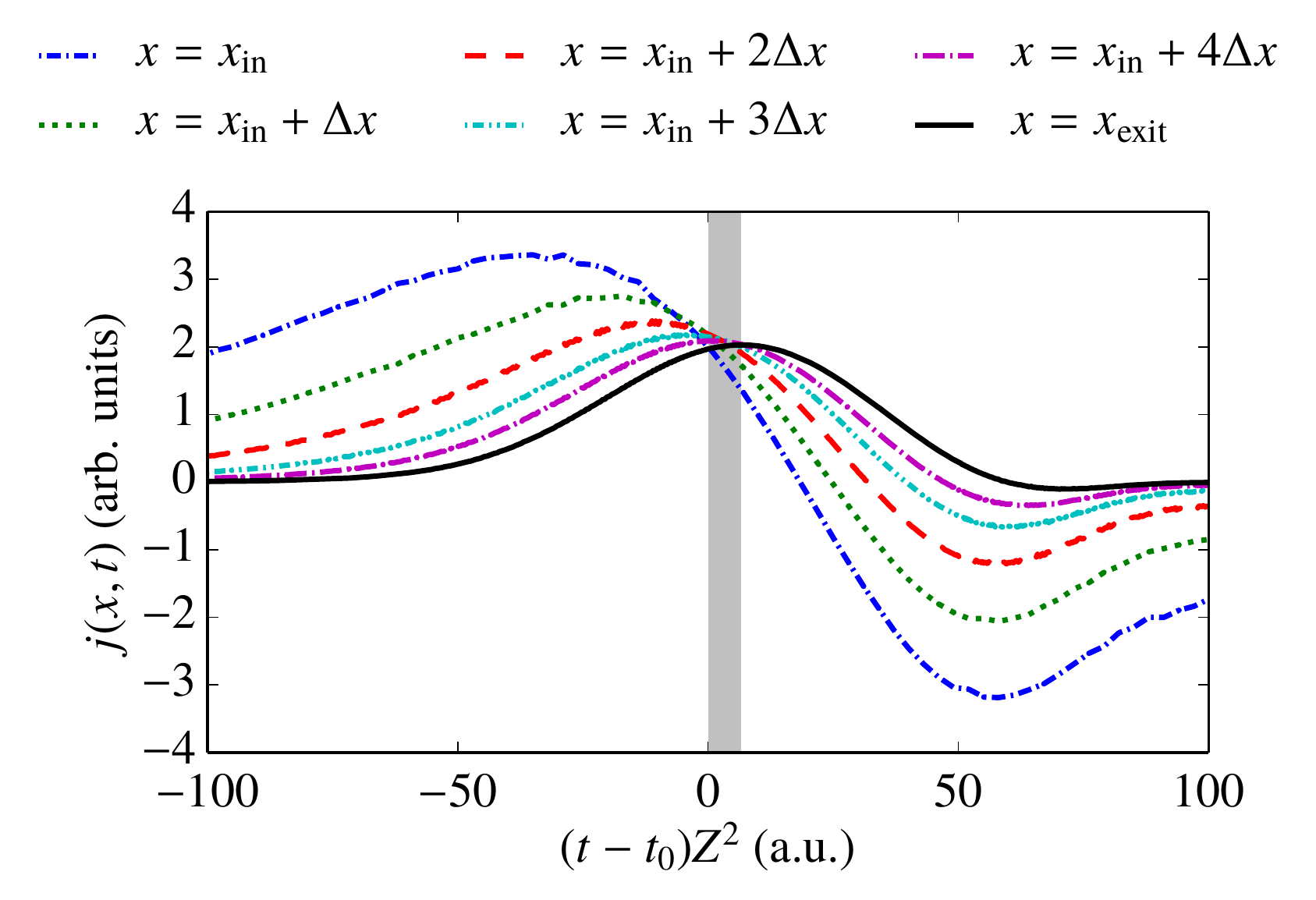}
  \caption{%(color online)
    The probability current $j(x,t)$ as a function of time at different
    positions $x$ between the barrier entry $x=x_\mathrm{in}$ and the
    barrier exit $x_\mathrm{exit}$ separated by $\Delta
    x=(x_\mathrm{exit}-x_\mathrm{in})/5$ for the parameters
    $E_0/Z^3=0.048$ and $\gamma=0.25$. The shadowed area represents the
    time between the instant of maximum electric field strength $t_0$
    and the instant of maximum probability at $x_\mathrm{exit}$.}
  \label{fig:many_flow}
\end{figure}

In experiments, a Coulomb-bound electron is usually excited by a laser
pulse with a wavelength much bigger than the atomic dimensions such that
the laser pulse is nearly homogeneous over the size of the atom.
Furthermore, relativistic effects and effects due to the magnetic field
component set in only for tunneling from highly charged ions
\cite{Yakaboylu:Klaiber:etal:2013:Relativistic_features}.  Thus, we will
apply the electric dipole approximation.  The laser pulse is modeled by
a time-varying homogeneous electric field
$E(t)=E_0\exp(-\omega^2(t-t_0)^2/2)$, where $t_0$ denotes the instant of
the maximum field strength $E_0$ and $\tau_\mathrm{E}=\sqrt2/\omega$ is
the time scale of the rise and decay of the electric field. The linear
polarization of the electric field and neglecting the magnetic field
component render the motion of the electron quasi-one-dimensional
allowing us to investigate general features of tunneling times in a
one-dimensional scenario.  Furthermore, tunneling in the fully
three-dimensional Coulomb problem can be described by an effective
one-dimensional tunneling barrier via introducing parabolic coordinates
\cite{Landau_3}.  Thus, we restrict ourselves to one-dimensional systems
and consider an electron bound to the soft-core potential
$-Z/\sqrt{x^2+\alpha(Z)}$
\cite{Hall:Saad:etal:2009:Energies_and,Clark:1997:Closed-form_solutions,Liu:Clark:1992:Closed-form_solutions,Su:Eberly:1991:Model_atom}
to model the essential features of an electron in a three-dimensional
Coulomb potential.  Here, $Z$ is the atomic number, and the softening
parameter $\alpha(Z)=2/Z^2$ is chosen such that the ground state energy
of the soft-core potential is $-I_\mathrm{p}=-Z^2/2$, which equals the
ground state energy of the Coulomb potential
\cite{weinberg2012lectures}.  Thus, the Schr{\"o}dinger equation (in atomic
units) \pagebreak[1]
\begin{equation}
  \label{eq:hamiltonian}
  \I\frac{\partial\psi}{\partial t} = \hat{H}(t) \psi =
  \left( 
    -\frac{1}{2}\frac{\partial^2}{\partial x^2} -
    \frac{Z}{\sqrt{x^2+\alpha(Z)}}-E(t)x 
  \right)\psi
\end{equation}
with the Hamiltonian $\hat H(t)$ and the effective potential $V(x,
t)=-{Z}/{\sqrt{x^2+\alpha(Z)}}-E(t)x$ will be solved numerically
\cite{*[{The dependence on $Z$ can be removed by suitable
    transformations of space, time, and the electric field strength, see
  }] [{.}] Bauke:etal:2011:Relativistic_ionization_characteristics}.
The so-called Keldysh parameter $\gamma=\omega\sqrt{2I_\mathrm{p}}/E_0$
\cite{keldysh} characterizes the ionization process as dominated by
tunneling for $\gamma\ll1$ and by multiphoton ionization for
$\gamma\gg1$.  Thus, simulation parameters will be set such that
$\gamma<1$ in the following.

\paragraph{Time delay $\tau_\mathrm{A}$}
 
The time delay $\tau_\mathrm{A}$ is based on the time-dependent
ionization rate.  In our one-dimensional model the probability current
$j(x,t)=(\psi(x, t)^*\partial_x\psi(x, t)-\psi(x, t)\partial_x\psi(x,
t)^*)/(2\I)$ represents the average net number of electrons passing a
given point at a specific time.  Thus, we determine the ionization rate
via the probability current at the exit $x_\mathrm{exit}$ as a function
of time, where $x_\mathrm{exit}$ is defined by the maximum electric
field strength $E_0$ via $V(x_\mathrm{exit},t_0)=-I_\mathrm{p}$.
Monitoring the probability current at a fixed position is justified,
because the tunnel probability is maximal for $E(t)=E_0$ and it is
exponentially suppressed for lower electric fields. Furthermore, %
$E(t)=\mbox{$E_0(1- \Delta t^2/\tau_\mathrm{E}^2 + O(\Delta
  t^4/\tau_\mathrm{E}^4))$}$ for $\Delta t=t-t_0$ with $|\Delta
t|<\tau_\mathrm{E}=2\sqrt{I_\mathrm{P}}/(\gamma E_0)$.  Thus the
tunneling barrier does not change substantially if times close to $t_0$
are considered.  To further ensure that the calculated exit
$x_\mathrm{exit}$ is where the electron exits the barrier, we place a
virtual detector at different points between $x_\mathrm{in}$ and
$x_\mathrm{exit}$ and calculate $j(x,t)$ at each point as a function of
time. As shown in Fig.~\ref{fig:many_flow}, the probability current has
a positive peak as well as a negative one for $x<x_\mathrm{exit}$,
indicating the tunneling and reflection dynamics, respectively.  This
tunneling and reflection dynamics corresponds to under-the-barrier
dynamics.  As reflection is absent for $x\ge x_\mathrm{exit}$, the
particle leaves the barrier at $x_\mathrm{exit}$.

\begin{figure}
  \centering
  \includegraphics[scale=0.45]{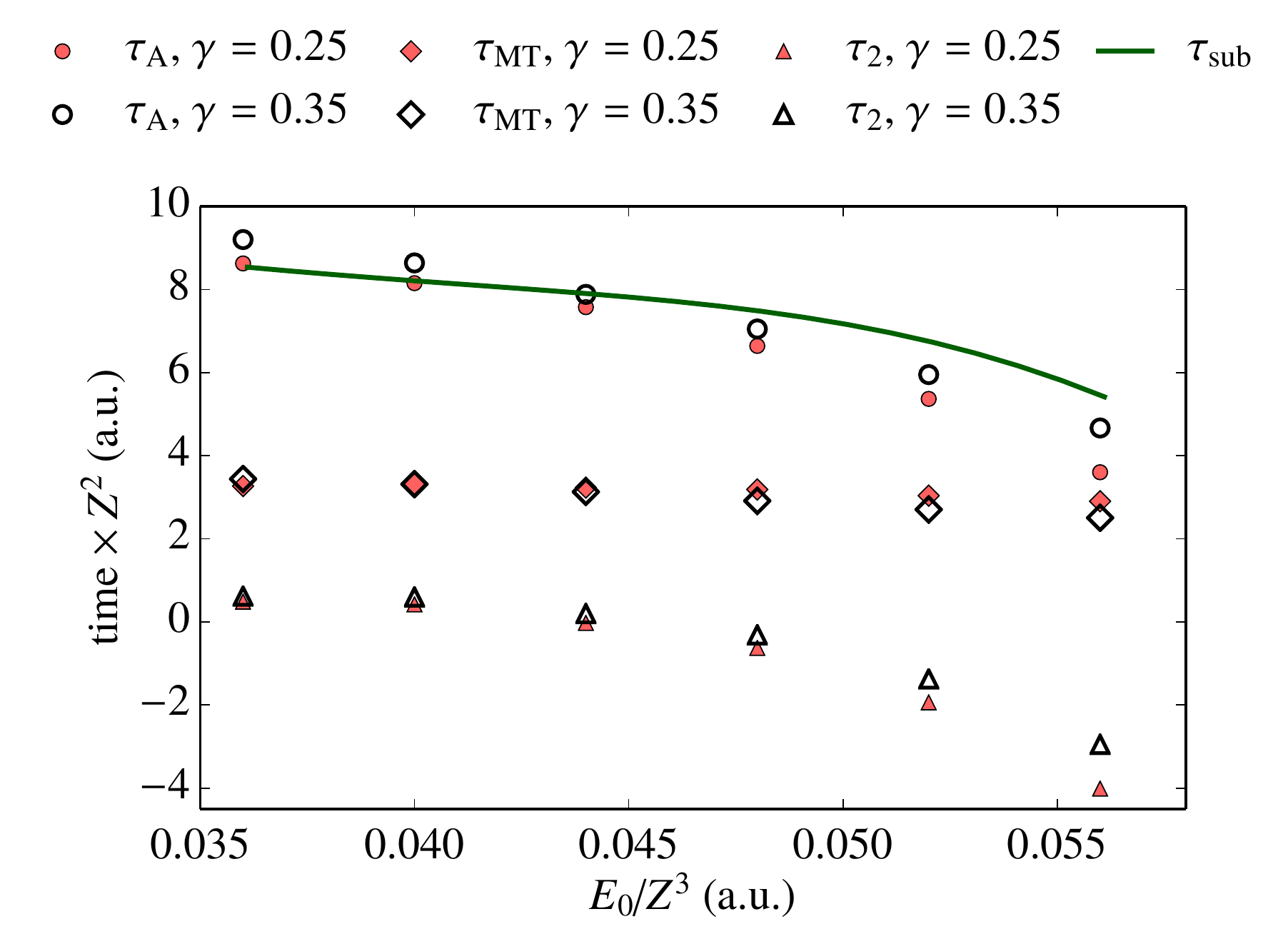}
  \caption{%(color online)
    The time delay $\tau_\mathrm{A}$, the Mandelstam-Tamm time
    $\tau_\mathrm{MT}$, and the times $\tau_2$ and $\tau_\mathrm{sub}$
    plotted for different electric field strengths $E_0$ and for
    different Keldysh parameters $\gamma$.  Definitions of the various
    times are given in the main text.}
  \label{fig:Atto}
\end{figure}

The most probable time delay $\tau_\mathrm{A}$ is calculated by
subtracting the instant of maximum field strength from the instant of
the maximum current, i.\,e., $\tau_\mathrm{A}=\argmax
j(x_\mathrm{exit},t)-\argmax E(t)$, which yields the positive time delay
shown in Fig.~\ref{fig:Atto}.  Note that, for the parameters used in
Fig.~\ref{fig:Atto}, $\tau_\mathrm{A}<\tau_\mathrm{E}$, and thus the
electric field remains almost constant for times $|\Delta
t|\lessapprox\tau_\mathrm{A}$, justifying our choice for
$x_\mathrm{exit}$.

The origin of the time delay $\tau_\mathrm{A}$ can be understood by
considering the time-energy uncertainty principle and following its
interpretation given by Mandelstam and Tamm
\cite{Mandelstam,Muga:etal:2007:Time,McDonald:Orlando:etal:2013:Tunnel_Ionization,Orlando:McDonald:etal:2014:Tunnelling_time}.
As a consequence of the time-energy uncertainty principle, the time,
which a wave function $\psi$ of a system with a time-independent
Hamiltonian $\hat{H}$ needs to change significantly, is bounded from
below by $1/(2\sigma_{\hat H})$, where $\sigma_{\hat
  H}=\sqrt{\vphantom{H_0^2}\smash{\braket{\psi|\hat{H}^2|\psi}-\braket{\psi|\hat{H}|\psi}^2}}$.
As outlined in the previous section, the Hamiltonian in
(\ref{eq:hamiltonian}) can be considered as time-independent for times
$|\Delta t|<\tau_\mathrm{E}$.  This allows us to define the
Mandelstam-Tamm time \pagebreak[2]
% \begin{equation}
%   \label{eq:tau_MT}
%   \tau_\mathrm{MT} = 
%     \left(2\textstyle
%       \sqrt{\vphantom{H_0^2}\smash{\braket{\psi(t_0)|\hat
%             H(t_0)^2|\psi(t_0)}-\braket{\psi(t_0)|\hat
%             H(t_0)|\psi(t_0)}^2}}
%     \right)^{-1}
%     \,,
% \end{equation}
\begin{equation}
  \label{eq:tau_MT}
  \tau_\mathrm{MT} = \frac{1}{2}
  \left(
    \braket{\psi(t_0)|\hat H(t_0)^2|\psi(t_0)} -
    \braket{\psi(t_0)|\hat H(t_0)|\psi(t_0)}^2
  \right)^{-1/2}
  \,,
\end{equation}
which is indeed a lower bound to the time delay $\tau_\mathrm{A}$ as
indicated in Fig.~\ref{fig:Atto}.  The time delay $\tau_\mathrm{A}$ is
close to its lower bound $\tau_\mathrm{MT}$, which indicates that the
delay $\tau_\mathrm{A}$ is a consequence of the wave function's inertia,
i.\,e., its inability to adopt instantaneously to the field.  The time
delay $\tau_\mathrm{A}$ decreases as $E_0$ increases at fixed Keldysh
parameter $\gamma$ and matches approximately $\tau_\mathrm{MT}$ when the
regime of over-the-barrier ionization is approached, which is for
$E_0/Z^3\approx 0.06\,\mathrm{a.u.}$

The observed decrease of the delay $\tau_\mathrm{A}$ with growing
electric field strength (but constant $\gamma$) is consistent with
calculations for the case of a sudden turn-on of the electric field,
which show that the time for the wave function to adopt to the electric
field is proportional to the Keldysh time
$\tau_\mathrm{K}=\sqrt{2I_\mathrm{p}}/E_0$
\cite{Orlando:McDonald:etal:2014:Tunnelling_time,McDonald:Orlando:etal:2013:Tunnel_Ionization},
although the functional dependence on $E_0$ is different for
continuously changing electric fields.  For a fixed maximum electric
field strength $E_0$ but increasing $\gamma$, the time delay
$\tau_\mathrm{A}$ increases.  As $\gamma$ increases, the pulse duration
decreases, granting the wave function less time $\sim\tau_\mathrm{E}$ to
evolve and to adopt to the time-varying Hamiltonian during the rise of
the electric field.  Thus, the wave function has less time to develop
the necessary components for tunneling and thus needs more time to reach
the maximum ionization rate.

The times $\tau_\mathrm{A}$ and $\tau_\mathrm{MT}$ are intractable by
analytical methods.  As we will demonstrate in the following, however,
one may employ the Wigner time as an approximation for
$\tau_\mathrm{A}$.  Comparing the quantum mechanical Wigner trajectory
$t_{\mathrm{W}}(x)$
\cite{Yakaboylu:Klaiber:etal:2014:Wigner_time,Yakaboylu:Klaiber:etal:2013:Relativistic_features}
to the trajectory $t_{\mathrm{c}}(x)$ of a classical particle, which
travels instantaneously from $x_\mathrm{in}$ to $x_\mathrm{exit}$ and
then moves according to the classical equations of motion, we can
calculate the time delay
\begin{equation}
  \label{eq:tau_sub}
  \tau_\mathrm{sub} = 
  t_{\mathrm{W}}(x_\mathrm{exit}) - t_{\mathrm{c}}(x_\mathrm{exit})\,,
\end{equation}
where the Wigner and the classical trajectories $t_{\mathrm{W}}(x)$ and
$t_{\mathrm{c}}(x)$ are determined such that both coincide at the entry
point $x_\mathrm{in}$.  

The delay $\tau_\mathrm{A}$ describes the time interval that the
electron spends under the barrier \emph{after the field maximum}.
Interpreting $\tau_\mathrm{A}$ as a tunneling time
\cite{Landsman:Weger:etal:2014:Ultrafast_resolution} is justified only
if the classical forbidden region is entered at the field maximum.  The
Wigner delay \eqref{eq:tau_sub}, however, represents the total time
spent under the barrier, since the relation
$t_{\mathrm{c}}(x_\mathrm{in}) = t_{\mathrm{c}}(x_\mathrm{exit})$ holds
in the classical model.  Also, the time delay $\tau_\mathrm{A}$ is often
understood as the time under the barrier.  Our numerical results show
that $\tau_\mathrm{sub}$ is indeed close to the time delay
$\tau_\mathrm{A}$, as shown in Fig.~\ref{fig:Atto}, where
$\tau_\mathrm{sub}$ is calculated for the one-dimensional Coulomb
potential.  Although $\tau_\mathrm{sub}$ and $\tau_\mathrm{A}$ agree
well, it is open if $\tau_\mathrm{A}$ should be interpreted as time
spent under the barrier% or just as a response time
% \cite{Orlando:McDonald:etal:2014:Tunnelling_time,McDonald:Orlando:etal:2013:Tunnel_Ionization}
, because the interpretation as under-the-barrier time is based on the
assumption that tunneling starts at the instant of the maximal electric
field, but the wave function could penetrate the tunneling barrier
earlier.

\begin{figure}
  \centering
  \includegraphics[scale=0.45]{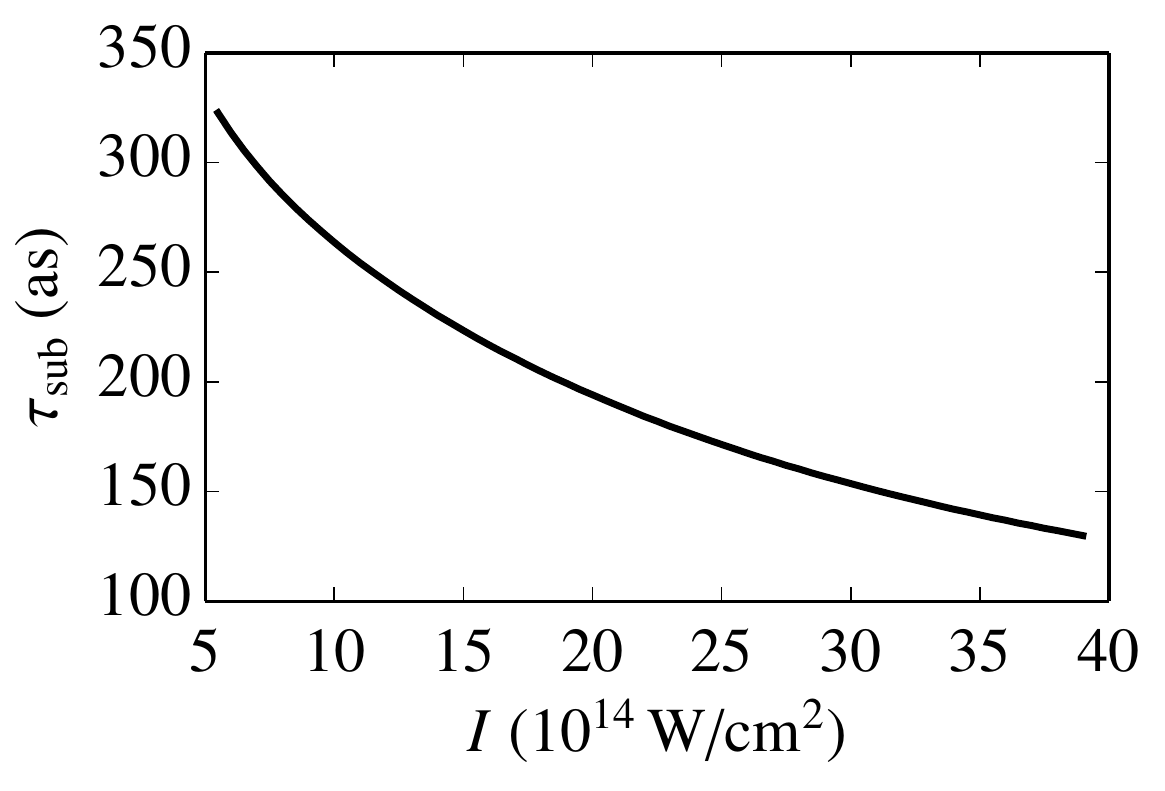}
  \caption{Wigner time $\tau_\mathrm{sub}$, which is an estimate for
    $\tau_{A}$, for the three-dimensional Coulomb potential including
    Stark shift and multielectron effects as a function of the
    electromagnetic field's peak intensity $I$ for ionization of helium
    atoms by linearly polarized laser fields.  The applied wavelength is
    $\lambda=800\,\mathrm{nm}$ and the Keldysh parameter varies from 0.6
    to 0.2.}
  \label{fig:tsub}
\end{figure}

For the one-dimensional Coulomb potential and at field intensities
approaching the over-the-barrier threshold, the Wigner time is given by
$\tau_\mathrm{sub}\approx14.29\sqrt{1-16E_0/Z^3}/Z^2$, and by
$\tau_\mathrm{sub}\approx9.0\sqrt{1-9.5E_0/Z^3}/Z^2$ for the
three-dimensional case.  Furthermore, the Wigner formalism can be
extended~\cite{Yakaboylu:etal} to include the Stark shift and also
multielectron effects.  The resulting Wigner time $\tau_\mathrm{sub}$
for helium is shown exemplarily in Fig.~\ref{fig:tsub} as a function of
the laser's peak intensity.

\paragraph{Exit momentum}

\begin{figure}
  \centering
  \includegraphics[scale=0.45]{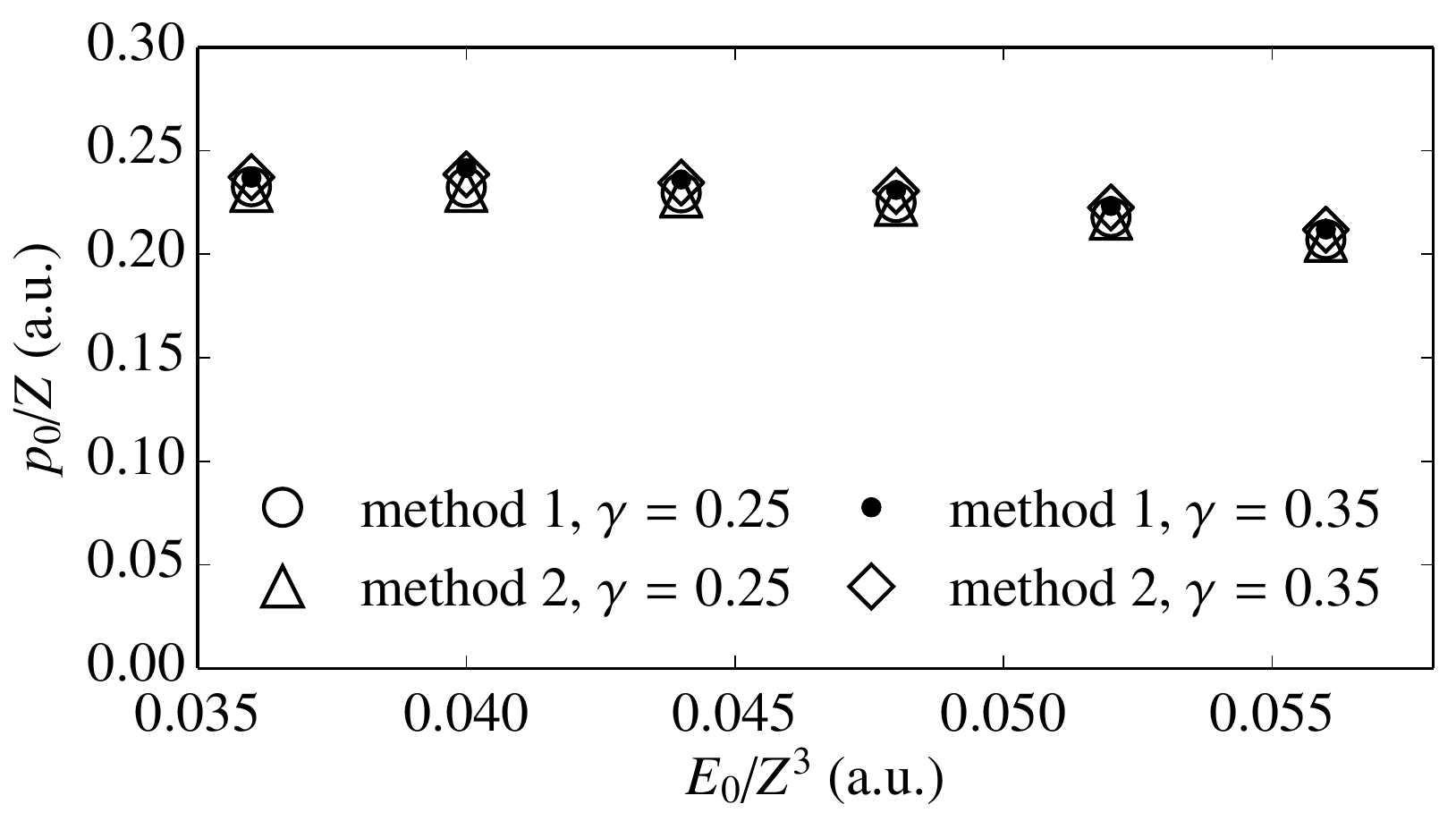}
  \caption{The momentum $p_0$ at the time $t_0+\tau_\mathrm{A}$ at the
    tunnel exit $x_\mathrm{exit}$ for different electric field strengths
    $E_0$ and different Keldysh parameters $\gamma$ as determined by two
    different methods.  Method~1 is based on the space-resolved momentum
    distribution, while method~2 utilizes the velocity of the
    probability flow; see the main text for details on both methods.}
  \label{fig:momentum}
\end{figure}

In the following, we determine the quantum mechanical exit momentum
$p_0$ in the direction of the electric field at the tunnel exit
$x_\mathrm{exit}$ at the instant of ionization $t=t_0+\tau_\mathrm{A}$
by two different methods.  In the first method, we calculate a
space-resolved momentum distribution around $x=x_\mathrm{exit}$ by
multiplying the wave function $\psi(x, t_0+\tau_\mathrm{A})$ by a
Gaussian window function with mean $x_\mathrm{exit}$ and width $\delta
x=(x_\mathrm{exit}-x_\mathrm{in})/20$ and calculating its Fourier
transform $\tilde\psi_\mathrm{exit}(t_0+\tau_\mathrm{A})$.  The most
probable exit momentum $p_0$ can be inferred by the momentum where
$|\tilde\psi_\mathrm{exit}(t_0+\tau_\mathrm{A})|^2$ is maximal.  In the
second method, the most probable initial momentum $p_0$ is determined
from the probability current $j(x_\mathrm{exit},t_0+\tau_\mathrm{A})$ at
the exit at the instant of the maximum probability current.  Following
Refs.~\cite{Wang:Tian:etal:2013:Extended_Virtual,Feuerstein:Thumm:2003:On_computation},
the local velocity of the wave function's probability flow at
$x_\mathrm{exit}$ equals
$v=j(x_\mathrm{exit},t_0+\tau_\mathrm{A})/|\psi(x_\mathrm{exit},t_0+\tau_\mathrm{A}
)|^2$.  Both methods yield very similar results for the moment $p_0$ as
shown in Fig.~\ref{fig:momentum}.  The initial momentum $p_0$ is almost
independent of the parameter $\gamma$ and depends only weakly on the
electric field strength~$E_0$.

As $p_0$ does not depend on the parameters of the external electric
field it must result from the initial quantum state, i.\,e., the ground
state of the binding potential.  In fact, the ground state of the
employed soft-core potential has in momentum space a width of about
$0.38\times Z$, which is of the same order as $p_0$; see
Fig.~\ref{fig:momentum}.  The fact that $p_0$ scales with the width of
the ground state's momentum distribution and not with its mean (which is
zero) may be interpreted as momentum components, which propagate into
the ionization direction, being ionized preferably.

\paragraph{Implications}

The above considerations are relevant for every high-precision
laser-induced tunneling experiment. Here we discuss as an example
attoclock measurements due to their currently high attention.  Attoclock
experiments aim to determine the time delay $\tau_\mathrm{A}$ between
the instant of the electric field maximum and the instant of tunneling,
which is not directly accessible experimentally.  Instead, the
asymptotic momentum of the tunneled electron is measured.  As it depends
on the exit momentum and on the moment at which the electron starts to
propagate freely in the field, one can infer $\tau_\mathrm{A}$ from the
electron's asymptotic momentum provided that the exit momentum is known.
The delay $\tau_\mathrm{A}$ is commonly reconstructed by assuming zero
initial momentum in the electric field direction.  Our numerical
simulations and the virtual detector approach, however, indicated a
nonzero initial momentum in the direction of the electric field.  How
does the zero-initial-momentum assumption affect the reconstruction of
$\tau_\mathrm{A}$ from the asymptotic momentum?

To answer this, we determine the final momentum of the tunneled electron
by propagating the wave function $\psi(t)$ till some final time
$t_\mathrm{f}$ such that $t_\mathrm{f}-t_0\gg1/\omega$ and separate the
tunneled part $\psi_\mathrm{free}(t_\mathrm{f})$ from the bound part of
the wave function by projecting out all bound eigenstates of $\hat{H}$
in Eq.~\eqref{eq:hamiltonian} for $E(t)=0$ from $\psi(t_\mathrm{f})$.
From the resulting probability density in momentum space representation
$\tilde\psi_\mathrm{free}(t_\mathrm{f})$ the most probable momentum
$p_\mathrm{f, q}$ can be inferred by the position where
$|\tilde\psi_\mathrm{free}(t_\mathrm{f})|^2$ is maximal. Using the
Newton equations of motion for an electron in the effective potential
$V(x, t)$, we can calculate at which instant of time the electron must
exit the barrier at $x_\mathrm{exit}$ with zero initial momentum in the
electric field direction such that its asymptotic momentum equals
$p_\mathrm{f, q}$. The result is the time delay $\tau_2$, which is also
shown in Fig.~\ref{fig:Atto} and that does \emph{not} coincide with the
delay $\tau_\mathrm{A}$.  The delay $\tau_2$ is close to zero or even
negative depending on the electric field strength as found in
Refs.~\cite{Torlina:Morales:etal:2015:Interpreting_attoclock,Yakaboylu:Klaiber:etal:2014:Wigner_time,Yakaboylu:Klaiber:etal:2013:Relativistic_features,Kaushal:Smirnova:2013:Nonadiabatic_Coulomb,Lein:2011:Streaking_analysis}.
Consequently, the delay $\tau_\mathrm{A}$ and the instant of tunneling
cannot be determined on the basis of the standard assumption of zero
initial momentum in the electric field direction. The exit momentum has
to be included.

Similarly to the delay $\tau_2$, also the time delays determined from
measured asymptotic momenta are close to zero (of the order of
experimental uncertainties)
\cite{Eckle:Pfeiffer:etal:2008:Attosecond_Ionization,Eckle:Smolarski:etal:2008:Attosecond_angular}.
As our numerical calculations indicate that the reconstruction of the
delay $\tau_\mathrm{A}$ from asymptotic momenta is sensitive to the
electron's exit momentum, we argue that a possible nonzero exit momentum
into the electric field direction has to be included for a reliable
reconstruction of $\tau_\mathrm{A}$ in attoclock experiments, not only a
nonzero exit momentum in the direction perpendicular to the electric
field as recently proposed
\cite{Klaiber:Hatsagortsyan:etal:2015:Tunneling_Dynamics,Kaushal:Smirnova:2013:Nonadiabatic_Coulomb}.

\paragraph{Conclusions}

We reexamined tunneling times in strong field ionization by an
\textit{ab~initio} solution of the time-dependent Schr{\"o}dinger equation.
Our calculations show that there is a delay $\tau_\mathrm{A}$ between
the maximum of the ionization rate and the maximum of the electric field
strength and a nonzero exit momentum $p_0$ in the electric field
direction.  The delay $\tau_\mathrm{A}$ can be explained as the response
time needed by the wave function to react to the change of the driving
electric field.  The initial momentum can be estimated from the width of
the ground state's momentum distribution, which is $Z$ for the
three-dimensional Coulomb potential.  The time $\tau_\mathrm{A}$ may be
estimated by the Wigner time $\tau_\mathrm{sub}$.  Note that our results
differ from the vanishing tunneling delay as obtained by the recently
introduced analytical $R$-matrix method
\cite{Torlina:Morales:etal:2015:Interpreting_attoclock}, which is based
on the assumption that this quasiclassical model provides a good
description of the quantum-mechanical tunneling dynamics also in the
vicinity of the tunneling barrier.

\acknowledgments The authors thank Karen~Z.\ Hatsgortsyan, Michael
Klaiber, Oleg Skoromnik, and Anton W{\"o}llert for valuable discussions.

\bibliography{attoclock_time}

\end{document}